\documentstyle[prd,aps,epsfig,floats]{revtex}
\begin{document}
\wideabs{
\title{Least square fitting with one parameter less}

\author{Bernd A. Berg}

\address{Department of Physics, Florida State University, 
         Tallahassee, FL 32306-4350} 

\date{\today } 

\maketitle
\begin{abstract}
It is shown that whenever the multiplicative normalization of a fitting
function is not known, least square fitting by $\chi^2$ minimization 
can be performed with one parameter less than usual by converting the 
normalization parameter into a function of the remaining parameters
and the data. 
\end{abstract} 
}
\narrowtext

{\bf Erratum:} Jochen Heitger and Johannes Voss of M\"unster University
found a typo in two subroutines which are used for the 4-parameter fit 
example given in subsection~\ref{subsec} of this paper. In each of the
subroutines,  \vskip 1.5pt

\centerline{\tt subg\_power2n.f and suby\_power2n.f,} \smallskip

\noindent {\tt a(2)} has to be replaced by {\tt a(1)} in the expression 
for the derivative {\tt dyda(1)}. Elimination of this error improves
the performance of the fitting method considerable.

In the following subsection~\ref{subsec} is re-written to reflect use 
of the correct subroutines and the error prone routines have been 
replaced in the package {\tt FITM1.tgz}, which is still available on 
the Web as described. Other parts of the paper are not affected.

\section{Introduction} \label{sec_intro}

The general situation of fitting by $\chi^2$ minimization is that $m$ 
data points $y_i=y(x_i)$ with error bars $\triangle y_i$ and a function 
$y(x;a_j)$ with $j=1,\dots,n$ parameters are given and we want to minimize
\begin{eqnarray} \label{chi2y}
  \chi^2 &=& \sum_{i=1}^m \left(\frac{y(x_i;a_j)-y_i}{\triangle y_i}
  \right)^2\
\end{eqnarray}
with respect to the $n$ parameters, where we neglect (as usual) 
fluctuations of the $\triangle y_i$ error bars.

In many practical application one of the $n$ parameters, say $a_n$, is 
the multiplicative normalization of the function $y(x;a_j)$ so that it 
can be written as 
\begin{eqnarray} 
  y(x;a_j) &=& a_n\,f(x;a_1,\dots,a_{n-1})\,.
\end{eqnarray}
With the parameters $a_1,\dots,a_{n-1}$ fixed there is a unique 
analytical solution $c_0$ for $c=a_n$, which minimizes $\chi^2(c)$, 
so that the function $y(x;a_j)$ depends effectively only on $n-1$ 
parameters:
\begin{eqnarray} \label{ym1}
  &y(x;a_1,\dots,a_{n-1})&\ =  \\ \nonumber
  &c_0(a_1,\dots,a_{n-1})&\,f(x;a_1,\dots,a_{n-1})\,.
\end{eqnarray}
This can be exploited to perform least square fitting with one parameter 
less. Remarkably, the number of steps required for one calculation of
$\chi^2$ remains linear in the number of data. Although the derivation 
of these result is rather straightforward, I have not encountered them
in the literature, though I found myself frequently in situations of 
performing least square fits of the type to which this method can be 
applied. 

In section~\ref{sec_const} explicit equations for $c_0(a_1, \dots, 
a_{n-1})$ and its derivatives with respect to the parameters are
derived. Practical examples based on Levenberg-Marquardt fitting 
\cite{Le44,Ma63} are given in section~\ref{sec_examples}. The 
conclusion from section~\ref{sec_examples} is that we have an 
additional useful approach, which mostly converges faster than the 
corresponding fit with the full number of parameters. Conclusion and 
an outlook on an eventual application are given in 
section~\ref{sec_summary}.  All examples of this paper can be 
reproduced with Fortran code that is provided on the Web and documented 
in appendix~\ref{sec_appendix}. In any case, the code provided here
should be useful.

\section{Calculation of the normalization constant and its derivatives} 
\label{sec_const}

We find the normalization constant $c_0(a_1,\dots,a_{n-1})$ by 
minimizing $\chi^2(c)$, $c=a_n$ for fixed parameters $a_1,\dots,
a_{n-1}$. We define
\begin{eqnarray}
  \chi^2(c) &=& \sum_{i=1}^m, \left(\frac{c\,f(x_i)-y_i}{\triangle y_i}
  \right)^2\,,
\end{eqnarray}
so that the derivative with respect to $c$ is zero at the minimum
$\chi^2(c_0)$:
\begin{eqnarray}
  0\ =\ \left.\frac{d\chi^2}{d\,c\,}\right|_{c=c_0}\ = \sum_{i=1}^m
  \frac{2\,f(x_i)\,\left(c_0\,f(x_i)-y_i\right)}{\left(\triangle 
  y_i\right)^2}\,.
\end{eqnarray}
This implies for $c_0$ the solution
\begin{eqnarray} \label{c0}
  c_0\ =\ \frac{r}{s}~~{\rm with}~~r &=& \sum_{i=1}^m r_i\,y_i\,,
  ~~r_i\ =\ \frac{f(x_i)}{\left(\triangle y_i\right)^2} \\ 
  \label{s} {\rm and}~~ s &=& 
  \sum_{j=1}^m \frac{f(x_j)^2}{\left(\triangle y_j\right)^2}\,.
\end{eqnarray}
As it should, this equation reduces for just one data point ($m=1$)
to $c_0=y_1/f(x_1)$. 

For fixed parameters $a_1,\dots,a_{n-1}$ the error bar $\triangle 
c_0$ of $c_0$ follows from the variances of the data points:
\begin{eqnarray} \label{c0va}
  \left.\left(\triangle c_0\right)^2\right|_a &=& s^{-2} 
  \sum_{i=1}^m \left(r_i\,\triangle y_i\right)^2\,,
\end{eqnarray}
where $|_a$ indicates that the parameters $a_1,\dots,a_{n-1}$ are 
kept fixed. When all error bars of the data agree, i.e., $\triangle 
y_i=\triangle y$ for $i=1,\dots,m$ holds, equations (\ref{c0}) and 
(\ref{c0va}) simplify to
\begin{eqnarray} 
  c_0 &=& \frac{\sum_{i=1}^m y_i\,f(x_i)}{\sum_{j=1}^m f(x_j)^2}\,, \\ 
  \left.\left(\triangle c_0\right)^2\right|_a &=&  
  (\triangle y)^2\,\frac{\sum_{i=1}^m f(x_i)^2}{\left( 
  \sum_{j=1}^m f(x_j)^2\right)^2}\,.
\end{eqnarray}
If, in addition, the function $f(x)$ is a constant, $f(x_i)=f_0$ for 
$i=1,\dots,m$, we find the usual reduction of the variance through 
sampling: $(\triangle c_0)^2|_a=(\triangle y/f_0)^2/m$. Note that 
the error bar Eq.~(\ref{c0va}) does not hold when the parameters 
$a_1,\dots,a_{n-1}$ are allowed to fluctuate, i.e., have themselves 
statistical errors $\triangle a_i$. Then the propagation of these 
errors into $c_0(a_1,\dots,a_{n-1})$ has to be taken into account, 
which is done below. Eq.~(\ref{c0va}) is mainly of relevance for the 
$n=1$ case when $c_0$ eliminates the sole parameter $a_1$.

In our illustrations based on the Levenberg-Marquardt approach as well 
as for many other fitting methods one needs the derivatives of the 
fitting function with respect to the parameters. With Eq.~(\ref{ym1}) 
this become
\begin{eqnarray} 
  \frac{dy}{da_j}=\frac{dc_0}{da_j}\,f+c_0\,\frac{df}{da_j}\,,~~~
  j=1,\dots,n-1\ .
\end{eqnarray}
We find the derivatives of $c_0$ from Eq.~\ref{c0}):
\begin{eqnarray} \label{dcda}
  \frac{dc_0}{da_j} &=& s^{-2}\,\left(s\,\frac{dr}{da_j} - 
  r\,\frac{ds}{da_j}\right)\,, \\
  \frac{dr}{da_j} &=& \sum_{i=1}^m \frac{y_i}{(\triangle y_i)^2}\,
  \frac{df(x_i)}{da_j}\,, \\
  \frac{ds}{da_j} &=& \sum_{i=1}^m \frac{2f(x_i)}{(\triangle y_i)^2}\,
  \frac{df(x_i)}{da_j}\,.
\end{eqnarray}
Using the derivatives (\ref{dcda}) the full variance of $c_0$ with the 
associated error bar $\triangle c_0=\sqrt{(\triangle c_0)^2}$ becomes
\begin{eqnarray} \label{c0v}
  \left(\triangle c_0\right)^2 &=& \left.\left(\triangle c_0\right)^2
  \right|_a + \sum_{j=1}^{n-1} \sum_{k=1}^{n-1}
  \frac{dc_0}{da_j}\,C_{jk}\,\frac{dc_0}{da_k}\,,
\end{eqnarray}
where $C_{jk}$ is the covariance matrix of the parameters 
$a_1,\dots,a_{n-1}$, which is in our examples returned by the
Levenberg-Marquardt fitting procedure.. 

\section{Examples} \label{sec_examples}

In this section we summarize results from least square fits with $n=1$ 
to $n=4$ parameters $a_i$, $i=1,\dots,n$, where the last one, $a_n$, 
is always taken to be a multiplicative normalization. The corresponding 
Fortran code is explained in appendix~\ref{sec_appendix}. 

We apply the Levenberg-Marquardt method in each case to all $n$ 
parameters as well as to $n-$1 parameters by considering $c_0$, the 
least square minimum of $a_n$ given by (\ref{c0}), to be part of the 
function. The Levenberg-Marquardt method uses steepest decent far from 
the minimum and switches to the Hessian approximation when the minimum 
is approached. Our Fortran implementation is a variant of the one of 
Ref.~\cite{Be04}. Besides the fitting function $y(x;a_i)$ one has to 
provide the derivatives 
\begin{eqnarray} \label{dyda}
  \frac{dy}{da_j}
\end{eqnarray}
and start values for the $n$, respectively $n-$1, parameters $a_i$. 
Usually the method will converge to the nearest local minimum of 
$\chi^2$, i.e., the minimum which has the initial parameters in its 
valley of attraction. 

Our choice of data for which we illustrate the method is rather 
arbitrary and emerged from considerations of convenience.
For the $n=1$ to $n=3$  parameter fits we use deconfining temperatures 
estimates from Markov Chain Monte Carlo (MCMC) simulations of 4D SU(2) 
gauge theory on $N_s^3N_{\tau}$ lattices as reported in Table~4 of 
a paper by Lucini et al.~\cite{Lu04}. We aim to extract from them 
corrections to asymptotic scaling by fitting $(aT_c)^{-1}=N_{\tau}
(\beta_c)$ to the form \cite{Ba06}
\begin{eqnarray} \label{flambda}
  y &=& N_{\tau}(\beta)\ =\ \frac{a_3}{f_{\lambda}(\beta)} \\
    &=& \frac{a_3}{f^{as}_{\lambda}(\beta,2)}\,\left(1+
        \frac{a_2}{\beta}+\frac{a_1}{\beta^2}\right)\,, \label{SU2fit}
\end{eqnarray}
where $N_{\tau}$ is the temporal extension of a $N_s^3N_{\tau}$ lattice
and $f^{as}_{\lambda}(\beta,N)$ is the universal two-loop asymptotic 
scaling function of SU(N) gauge theory
\begin{eqnarray} 
  f^{as}_{\lambda}(\beta,N)\ =\ e^{-1/(2b_0g^2)}\,
  (b_0g^2)^{-b_1/(2b_0^2)},\ g^2=\frac{2N}{\beta}
\end{eqnarray}
with $b_0=(N/16\pi^2)\,(11/3)$ the one-loop \cite{Gr73,Po73} and
$b_1=(N/16\pi^2)^2 (34/3)$ the two-loop result \cite{Jo74,Ca74}.

In Table~4 of \cite{Lu04} the error bars are for the critical coupling 
constants $\beta_c$. For the purpose of the fit (\ref{SU2fit}) the 
error bars are shuffled to $N(\beta_c)$ by means of the equation
\begin{eqnarray} 
  \triangle (aT_c)^{-1} &=& \frac{N_{\tau}}{f_{\lambda}(\beta_c)}\,
  \left[f_{\lambda}(\beta_c)-f_{\lambda}(\beta_c+\triangle\beta_c)
  \right]\,,
\end{eqnarray}
where a preliminary estimate of the $f_{\lambda} (\beta)$ scaling 
function (\ref{flambda}) is used. The thus obtained data (omitting 
$N_{\tau}<4$ lattices) are compiled in Table~\ref{tab_data}.

\begin{table}[ht] 
\centering
\caption{\label{tab_data}{Data used for our least square fitting examples.}}
\smallskip
\begin{tabular}{|c|c|c|c|}
\multicolumn{2}{|c|}{Lucini et al.\ \cite{Lu04}}&
\multicolumn{2}{c|}{Bhanot et al. \cite{Bh87}}\\ \hline
$\beta_c$ & $N_{\tau}(\beta_c)$ & $N_s$ & Im$(u)$  \\ \hline
2.29860   & 4.0000 (77)         & ~4    & 0.087739 (5) \\
2.37136   & 5.0000 (86)         & ~5    & 0.060978 (5) \\
2.42710   & 6.0000 (32)         & ~6    & 0.045411 (5) \\
2.50900   & 8.0000 (32)         & ~8    & 0.028596 (5) \\
 $-$      &  $-$                & 10    & 0.019996 (5) \\
\end{tabular} \end{table} 

For the $n=4$ parameter fits results from Bhanot et al.~\cite{Bh87}
for the imaginary part Im$(u)$ of the partition function zero closest 
to the real axis are used, which are obtained from MCMC simulation of 
the 3D Ising model on $N_s^3$ lattices. These data are also collected 
in our Table~\ref{tab_data}.

To leading order their finite size behavior is
\begin{eqnarray} \label{nu2par}
  y\ =\ {\rm Im}(u)\ =\ a_2\,x^{a_1}~~{\rm with}~~x=N_s\,,
\end{eqnarray}
where $a_1$ is related to the critical exponent $\nu$ of the correlation
length by $a_1=-1/\nu$. In the context of our method this 2-parameter 
fit is of no interest, because it can be mapped onto linear regression
for which the $\chi^2$ minimum leads to analytical solutions for both 
parameters, $a_1$ and $a_2$. For the critical exponent $\nu$ this yields 
$1/\nu = 1.6185 (2)$, but has an unacceptable large $\chi^2$, which 
leads to $Q=0$ for the goodness of fit (details are given in 
\cite{Be04}).

One is therefore led to including subleading corrections by moving to
the 4-parameter fit
\begin{eqnarray} \label{nu4par}
  y\ =\ {\rm Im}(u)\ =\ a_4\,x^{a_1} \left(1+a_2x^{a_3}\right)\ 
     =\ a_4\,f(x)
\end{eqnarray}
for which our method replaces $a_4$ by $c_0$ (\ref{c0}) as function 
of the other parameters. We are now ready to present the results for
our fits.

\subsection{1-parameter fits}

The function (\ref{flambda}) reduces to the from
\begin{eqnarray} \label{fla1}
  y &=& N_{\tau}(\beta)\ =\ \frac{a_1}{f^{as}_{\lambda}(\beta)}
\end{eqnarray}
and there are no fitting parameters left when the analytical solution 
$c_0$ (\ref{c0}) with the error bar $\triangle c_0$ form (\ref{c0va}) 
is used for $a_1$.

Our Levenberg-Marquardt procedure works down to a single fit parameter 
and uses besides the fitting function (\ref{fla1}) the only derivative
\begin{eqnarray} \label{dyda1}
  \frac{dy}{da_1} &=& \frac{1}{f^{as}_{\lambda}(\beta)}\,. 
\end{eqnarray}
Using the start value $a_1=0.0628450$ one finds convergence after three
iteration with the results
\begin{eqnarray} \label{1par}
  a_1\ =\ 0.025336\ (26)~~{\rm and}~~\chi^2\ =\ 4263\,.
\end{eqnarray}
Without any iteration identical values for $a_1$ and $\chi^2$ are 
obtained by using the analytical solution $c_0$ and its error bar 
(\ref{c0va}). Note that $\chi^2$ has to be the same for identical 
parameters. So $c_0$ still counts when it comes to counting the 
degrees of freedom.

\begin{figure}[tb] \begin{center} %
\epsfig{figure=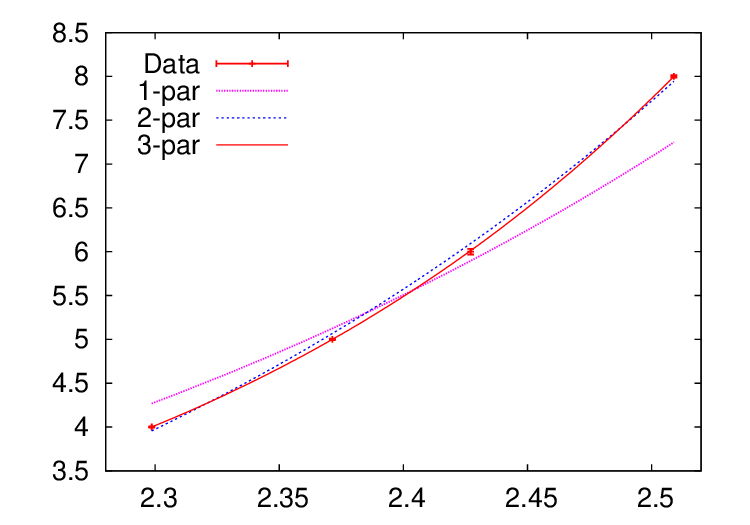,width=\columnwidth} 
\caption{Uniform distribution: Impossible (left) and missing events 
(in the enlarged thickness of the right border).  \label{fig_Tc}} 
\end{center} \end{figure}

Obviously the obtained $\chi^2$ is unacceptable large and it is well 
visible from the 1-par curve in Fig.~\ref{fig_Tc} that this fit is not 
good. Additional parameters are needed to account for corrections to 
asymptotic scaling.

\subsection{2-parameter fits}

The function (\ref{flambda}) is now reduced to the form
\begin{eqnarray} \label{fla2}
  y &=& N_{\tau}(\beta)\ =\ \frac{a_2}{f^{as}_{\lambda}(\beta)}\,
  \left(1+\frac{a_1}{\beta}\right)\,. 
\end{eqnarray}
For the fit with two parameters we use the start values $a_2=
0.0628450$ (as for $a_1$ before) and $a_1=-1.43424$. After six 
iterations we find convergence and the results
\begin{eqnarray} \label{2par}
  a_1 = -1.5751\,(84)\,,\ a_2 = 0.07460\,(77)\,,\ \chi^2 = 117\,.
\end{eqnarray}
Running the fitting routine now just for the parameter $a_1$ by 
eliminating $a_2$ in the described way, we find the results (\ref{2par}) 
after four iterations. In particular, also the error bar of $c_0$, now 
calculated via Eq.~(\ref{c0v}), agrees with the error bar of $a_2$.

Clearly the $\chi^2$ is still too large to claim consistency between
the fit and the data though the visible improvement is considerable.
See the 2-par curve in Fig.~\ref{fig_Tc}.

\subsection{3-parameter fits}

We fit now to the full functional form (\ref{flambda}). For the three
parameter fit the previous starting values are re-shuffled $a_2\to a_3$,
$a_1\to a_2$ and the additional starting value is taken to be $a_1=1$.
Our Levenberg-Marquardt procedure needs 245 iterations for convergence
and yields the values
\begin{eqnarray} \label{3par1}
  a_1 &=& 4.685\ (83)\,,~~a_2\ =\ -4.199\ (46)\,,\\
  a_3 &=& 0.395\ (30)\,,~~\chi^2 = 0.156\,, \label{3par2}
\end{eqnarray}
which are rather different than the corresponding results $a_3\to a_2$
and $a_2\to a_1$ (\ref{2par}) of the 2-parameter fit.

Identical values (\ref{3par1}), (\ref{3par2}) are obtained after 
eliminating the normalization, here $a_3$, from the fit parameters 
used in the Levenberg-Marquardt iteration and we find a reduction of 
iterations from 245 to 12.

The $\chi^2$ is now small enough to signal consistency between the data
and fit, although the 2-par and 3-par curves in Fig.~\ref{fig_Tc} are
hard to distinguish visually. Converting the $\chi^2$ of (\ref{3par2})
into a goodness of fit one finds $Q=0.69$.

\subsection{4-parameter fits} \label{subsec}

The aim is to perform the 4-parameter fit (\ref{nu4par}) for the 
data of Bhanot et al.\ (Table~\ref{tab_data}). Using all four
parameters, our first set of starting values,
$$ a_1=-1.6\,,~~a_2=0.1\,,~~a_3=-1.0\,,~~a_4=0.8\,, $$
is chosen so that $a_1$ stays close to the 2-parameter fit result 
$-a_1=1/\nu=1.6185\,(2)$ from \cite{Be04}. Running our
Levenberg-Marquardt fit program on these initial values,
it needs 391 iterations to converge and finds
\begin{eqnarray} \label{4par1}
  a_1 &=&-1.5981\,(31)\,,~~a_2\ =\ 0.77\,(39)\,,\\ \label{4par2}
  a_3 &=&-2.80\,(52)\,,~~a_4\ =\ 0.7917\,(61)\,,~~\chi^2 = 0.113\,, 
\end{eqnarray}
where the $\chi^2$ can be converted into a goodness of fit $Q=0.74$. 

Eliminating the normalization $a_4$ from the direct fitting parameters,
convergence is reached after 58 iterations and we find identical
estimates as before:
\begin{eqnarray} \label{4m1par1}
  a_1 &=&-1.5981\,(31)\,,~~a_2\ =\ 0.77\,(39)\,,\\ \label{4m1par2}
  a_3 &=&-2.80\,(52)\,,~~c_0\ =\ 0.7917\,(61)\,,~~\chi^2 = 0.113\,, 
\end{eqnarray}
Noticeably, our example features a rugged landscape for $\chi^2$ as 
function of the parameters. With different starting values 
$$ a_1=-4.4\,,~~a_2=1.3\,,~~a_3=2.8\,,~~a_4=0.6 $$
a minimum entirely different from the one above was found by accident. 
Running our Levenberg-Marquardt procedure on these starting values
convergence is reached after 9 iterations and gives
\begin{eqnarray} \label{4par3}
  a_1&=&-4.40\,(53)\,,~~a_2\ =\ 1.31\,(66)\,,\\ \label{4par4}
  a_3&=& 2.80\,(52)\,,~~a_4\ =\ 0.61\,(31)\,,~~\chi^2 = 0.113\,. 
\end{eqnarray}
Eliminating the normalization $a_4$ we find convergence after 8
iterations and the same results:
\begin{eqnarray} \label{4m1par3}
  a_1&=&-4.40\,(53)\,,~~a_2\ =\ 1.31\,(66)\,,\\ \label{4m1par4}
  a_3&=& 2.80\,(52)\,,~~a_4\ =\ 0.61\,(31)\,,~~\chi^2 = 0.113\,. 
\end{eqnarray}
Either set of parameters fits the data perfectly well in their range,
while the different fits function diverge quickly out of this range, 
i.e., for larger lattices.

\section{Conclusion and Outlook} \label{sec_summary}

This paper shows that we can exclude the multiplicative normalization
of a fitting function from the variable parameters of a $\chi^2$ 
minimization and include it into the fitting function (it still 
counts when it comes to determining the degrees of freedom). Our 
simple examples show that this works well, reducing the number of 
iterations.

The code discussed in appendix~\ref{sec_appendix} shows that there is 
no extra work involved for the user. Once the general, application 
independent, code is set up, the subroutine which the user has to
supply has one parameter less than the one needed for the usual 
Levenberg-Marquardt fitting procedure (compare the reduction from 
{\tt subg\_la3su2.f} to {\tt suby\_la3su2.f}). Besides, it is 
useful to have alternatives at hand when one is trying to find 
convenient initial values.

Finally, there may be interesting applications, which cannot be easily
incorporated into conventional fitting schemes. For instance, there 
are situations where distinct data sets are supposed to be described 
by the same function with different multiplicative normalizations. An
example is scale setting in lattice gauge theories \cite{So13}. The
method of this paper can then be used to eliminate all multiplicative
constants from the independent parameters of the fit so that one can 
consolidate all data sets into one fit. This application will be pursued 
elsewhere.

\acknowledgments
This work was in part supported by the US Department of Energy under 
contract DE-FG02-13ER41942. I would like to thank Jochen Heitger and 
Johannes Voss for communicating the error which is corrected in 
this second version of the paper.

\appendix

\section{Fortran implementation} \label{sec_appendix}

For a limited time the Fortran code can still be downloaded as archive 
{\tt FITM1.tgz} from the authors website at\footnote{A previous version 
of this paper has been published in Computer Physics Communication 200 
(2016) 254-258 and the program package is available from their program 
library. When using these programs, the typo in the two relevant 
subroutines needs to be corrected to reproduce the results of 
subsection~\ref{subsec}.}

\noindent http://www.hep.fsu.edu/\~\,$\!$berg/research/research.html\,.

\noindent After downloading the {\tt FITM1.tgz} file, it unfolds under 
$$\rm tar\ -\!zxvf\ FITM1.tgz $$
into a tree structure with the folder {\tt FITM1} at its top. On 
the next level there are two subfolders {\tt examples} and {\tt libs}.
To reproduce the results of section~\ref{sec_examples} go to the
{\tt examples} folder, where one finds the subfolders {\tt 1par}, {\tt 
2par}, {\tt 3par} and {\tt 4par}. 

Each of these subfolders contains two Fortran programs, $n${\tt gfit.f} 
and $n${\tt gfitm1.f}, where $n=1,2,3$ or 4 denotes the number of 
parameters. Both programs are ready to be compiled and run, say
with {\tt ./a.out$>$a.txt}. The thus produced results should agree
with those found in the text file {\tt a$n$.txt} and {\tt a$n$m1.txt}.
In addition graphical output is created. Type
$$ \rm gnuplot\ gfit.plt $$
to see the plots (a gnuplot driver {\tt gfit.plt} is located in each 
folder).

The programs read data and starting values from files named {\tt 
fort.10}. For the $n=4$ case there are two sets, {\tt bhanot1.dat} and 
{\tt bhanot2.dat}, which differ by the initial values and the desired 
one has to be copied on {\tt fort.10} before the run.

With exception of 1gfitm1.f, which is a special case discussed at the
end of this appendix, the programs rely on our Levenberg-Marquardt 
subroutine {\tt fitsub} to find the minimum of $\chi^2$. This subroutine, 
is transferred after the {\tt end} statement of each main program by an 
{\tt include} statement into the code,
\begin{small} \begin{verbatim}
      include '../../libs/fortran/fitsub.f'
\end{verbatim} \end{small} \smallskip
and in the same way this is done for all others routines needed. These 
{\tt include} statements allow for easy tracking of the location of 
the source code of each routine.

The crucial difference between the two main programs is: $n${\tt gfit.f} 
calls {\tt fitsub} for {\tt nfit}=$n$ parameters, but $n${\tt gfitm1} 
calls it for {\tt nfit}$=$n$-1$ parameters. This is achieved by feeding 
distinct subroutines {\tt subg} into {\tt fitsub}, which define the 
fitting functions and their derivatives with respect to the {\tt nfit} 
parameters. For instance, for the 3-parameter case the subroutine used 
by the run of {\tt 3gfit.f} is {\tt subg\_la3su2.f} as listed here:
\begin{small} \begin{verbatim}
      subroutine subg(x,a,yfit,dyda,nfit) 
c BB May 20 2015. User provided subroutine
c for a 3-parameter fit of the pure su2 scale,
c yfit=a3*(1+a2/x+a1/x**2)/flasu2(x,2) with x=beta 
c and flasu2(x,2) the asymptotic su2 scaling 
c function (yfit has the dimension of a length).
      include '../../libs/fortran/implicit.sta'
      include '../../libs/fortran/constants.par'
      dimension a(nfit),dyda(nfit)
      if(nfit.ne.3) stop "subg_la3su2: nfit.ne.3."
      x2=x**2
      flasu2=fla(x,itwo)
      dyda(3)=(one+a(2)/x+a(1)/x2)/flasu2
      yfit=a(3)*dyda(3)
      dyda(2)=(a(3)/x)/flasu2
      dyda(1)=(a(3)/x2)/flasu2
      return
      end
\end{verbatim} \end{small} \smallskip
The {\tt 3gftim1.f} program and all other $n${\tt gfitm1.f} programs,
but $n$=1, include {\tt subgfitm1.f} with the following code:
\begin{small} \begin{verbatim}
      subroutine subg(xx,aa,yfit,dyda,nfit) 
c BB May 20 2015. Generic routine for least square 
c fit with one parameter less. Input needed: User 
c provided subroutine suby for unnormalized fit
c function and its derivatives with respect to 
c its parameters.
      include '../../libs/fortran/implicit.sta'
      include '../../libs/fortran/constants.par'
      include '../../libs/subs/common_fitting.f' 
c For this routine the common transfers x,y,ye.
      dimension aa(nfit),dyda(nfit)
      call chi2dcda(ndat,nfit,x,y,ye,aa,dcda,c0)
      call suby(xx,aa,yy,dyda,nfit)
      yfit=c0*yy
      do i=1,nfit
        dyda(i)=c0*dyda(i)+dcda(i)*yy
      enddo
      return
      end
\end{verbatim} \end{small} \smallskip
Here the call to {\tt suby} defines the user supplied function and 
its derivatives as the {\tt subg} routines do for the $n${\tt gfit.f} 
programs. The {\tt suby} routines replace the normalization parameters
by the number one. For our example above {\tt subg\_la2su2.f} becomes 
{\tt suby\_la2su2.f} given by
\begin{small} \begin{verbatim}
      subroutine suby(x,a,y,dyda,nfit) 
c BB may 20 2015. User provided fit function 
c y=(1+a2/x+a1/x**2)/flasu2(x,2) and derivatives.
      include '../../libs/fortran/implicit.sta'
      include '../../libs/fortran/constants.par'
      dimension a(nfit),dyda(nfit)
      if(nfit.ne.2) stop "suby_la3su2: nfit.ne.2."
      x2=x**2
      flasu2=fla(x,itwo)
      y=(one+a(2)/x+a(1)/x2)/flasu2
      dyda(2)=(one/x)/flasu2
      dyda(1)=(one/x2)/flasu2
      return
      end
\end{verbatim} \end{small} \smallskip
This it the only routine, which the user has to define for the
reduced fitting procedure. The {\tt chi2dcda} routine called in 
{\tt subgfitm1.f} is generic for all these fits and calculates 
$c_0$ (\ref{c0}) and its derivatives (\ref{dcda}) as defined 
in section~\ref{sec_const} with the following code:
\begin{small} \begin{verbatim}
      subroutine chi2dcda(n,nf,x,y,ye,a,dcda,c0) 
c BB May 20 2015. Determination of the normalization 
c constant c0 for chi2 fit of data y_i with the fit 
c function c0*f. Then, in dcda the derivatives of 
c c0 with respect to the fit parameters.
c Input: data arrays x(n),y(n),ye(n), 
c        parameter array a(nf).
c Output: dcda(nf), c0.
c        variable fit parameters a(nf).
      include '../../libs/fortran/implicit.sta'
      include '../../libs/fortran/constants.par'
      parameter(mf=30) ! maximum number parameters.
      dimension drda(mf),dsda(mf),dfda(mf),a(nf),
     &  dcda(nf), x(n),y(n),ye(n)
      if(nf.gt.mf) stop "chi2dcda: nf.gt.mf."
      r=zero
      s=zero
      do j=1,nf
        drda(j)=zero
        dsda(j)=zero
      enddo
      do i=1,n
        call suby(x(i),a,f,dfda,nf)
        ye2i=one/ye(i)**2 ! 1/(error bar squared).
        ri=f*ye2i
        r=r+y(i)*ri     ! iterate r.
        s=s+f*ri        ! iterate s.
        addy=y(i)*ye2i
        addf=two*ri
        do j=1,nf
          drda(j)=drda(j)+addy*dfda(j)
          dsda(j)=dsda(j)+addf*dfda(j)
        enddo
      enddo
      c0=r/s
      do j=1,nf
        dcda(j)=(s*drda(j)-r*dsda(j))/s**2
      enddo
      return
      end
\end{verbatim} \end{small} \smallskip
An extension of this routine, {\tt chi2c0eb.f}, calculates also the
variance (\ref{c0v}) of $c_0$ and needs on input the covariance matrix 
of the parameters $a_1,\dots,a_{n-1}$.

Finally, the program {\tt 1gfitm1.f} is a special case, because the
{\tt fitsub} routine does not work for {\tt nfit=0} parameters. 
Actually, it is not needed at all in this case and becomes replaced 
by a variant of the {\tt chi2dcda.f} routine: {\tt chi2c0.f} 
calculates $c_0$ (\ref{c0}) and its error bar, now according to
Eq.~(\ref{c0va}) instead of (\ref{c0v}).

\end{document}